\documentclass[
  floatfix,
  reprint,
  superscriptaddress,
  amsmath,amssymb,
  aps,
  prc,
]{revtex4-2}
\usepackage{amsmath,amssymb}
\usepackage{graphicx}
\usepackage{booktabs}
\usepackage{hyperref}
\usepackage{orcidlink}
\usepackage{multirow}

\begin{document}

\title{Measurement of the $^{169}$Tm$(\gamma,n)$$^{168}$Tm Cross Section and Implications for $\gamma$-Process Calculations}

\author{De-Xin Wang~\orcidlink{0000-0002-3980-9910}}
\thanks{These authors contributed equally to this work.}
\affiliation{Institute of Nuclear Physics, College of Physics and Electronic Information, Inner Mongolia Minzu University, Tongliao 028000, China}

\author{Su-Ya-La-Tu Zhang~\orcidlink{0000-0002-9580-4209}}
\email{zsylt0416@163.com}
\affiliation{Institute of Nuclear Physics, College of Physics and Electronic Information, Inner Mongolia Minzu University, Tongliao 028000, China}

\author{Zi-Rui Hao~\orcidlink{0000-0001-5538-4118}}
\email{haozr@sari.ac.cn}
\affiliation{Shanghai Advanced Research Institute, Chinese Academy of Sciences, Shanghai 201210, China}

\author{Han Guo~\orcidlink{0009-0005-3536-2874}}
\thanks{These authors contributed equally to this work.}
\affiliation{Institute of Nuclear Physics, College of Physics and Electronic Information, Inner Mongolia Minzu University, Tongliao 028000, China}

\author{Wen-Yu Xin~\orcidlink{0000-0003-3646-9356}}
\affiliation{Institute of Astronomy and Physics, Inner Mongolia University, Hohhot 010021, China}

\author{Guo Li~\orcidlink{0000-0002-3856-1108}}
\affiliation{Institute of Nuclear Physics, College of Physics and Electronic Information, Inner Mongolia Minzu University, Tongliao 028000, China}
\affiliation{CAS Key Laboratory of Optical Astronomy, National Astronomical Observatories, Chinese Academy of Sciences, Beijing 100101, China}

\author{Yong-Shun Huang~\orcidlink{0009-0008-5637-5342}}
\affiliation{Institute of Nuclear Physics, College of Physics and Electronic Information, Inner Mongolia Minzu University, Tongliao 028000, China}

\author{Jin-Feng Bai~\orcidlink{0009-0005-3906-8250}}
\affiliation{Institute of Nuclear Physics, College of Physics and Electronic Information, Inner Mongolia Minzu University, Tongliao 028000, China}

\author{Kai Wang~\orcidlink{0009-0002-1300-9752}}
\affiliation{Institute of Nuclear Physics, College of Physics and Electronic Information, Inner Mongolia Minzu University, Tongliao 028000, China}

\author{Mei-Rong Huang~\orcidlink{0000-0001-5870-5239}}
\affiliation{Institute of Nuclear Physics, College of Physics and Electronic Information, Inner Mongolia Minzu University, Tongliao 028000, China}

\author{Bao-Shan Xi~\orcidlink{0000-0002-3785-9113}}
\affiliation{Institute of Nuclear Physics, College of Physics and Electronic Information, Inner Mongolia Minzu University, Tongliao 028000, China}

\author{Peter Mohr~\orcidlink{0000-0002-6695-9359}}
\affiliation{HUN-REN Institute of Nuclear Research (ATOMKI), H-4001 Debrecen, Hungary}

\author{Jian-Fang Zhang~\orcidlink{0009-0004-1648-6639}}
\affiliation{Institute of Nuclear Physics, College of Physics and Electronic Information, Inner Mongolia Minzu University, Tongliao 028000, China}

\author{Gong-Tao Fan~\orcidlink{0000-0002-4240-7293}}
\affiliation{Shanghai Advanced Research Institute, Chinese Academy of Sciences, Shanghai 201210, China}
\affiliation{University of Chinese Academy of Sciences, Beijing 100049, China}

\author{Hong-Wei Wang~\orcidlink{0000-0002-8178-0605}}
\affiliation{Shanghai Advanced Research Institute, Chinese Academy of Sciences, Shanghai 201210, China}
\affiliation{University of Chinese Academy of Sciences, Beijing 100049, China}


\begin{abstract}
The $^{169}\mathrm{Tm}(\gamma,n)^{168}\mathrm{Tm}$ cross section has been measured at the Shanghai Laser Electron Gamma Source (SLEGS) using quasi-monoenergetic laser-Compton-scattering $\gamma$ rays over the energy range from the neutron threshold to 14.61~MeV.
The photon flux and energy distribution were determined with a LaBr$3$(Ce) spectrometer, and monoenergetic cross sections were obtained using an iterative unfolding method.
Below 12~MeV, the present cross sections agree well with the NewSUBARU data and the IAEA-2019 evaluations. However, near the maximum of the $(\gamma,n)$ excitation function, the present cross sections are significantly higher than both the NewSUBARU data and the IAEA-2019 evaluated values.
The $E1$ $\gamma$-ray strength functions were derived by fitting the experimental cross sections measured at SLEGS and NewSUBARU.
The stellar reaction rates for $^{169}\mathrm{Tm}(\gamma,n)$ and $^{168}\mathrm{Tm}(n,\gamma)$ were calculated by incorporating these constrained $\gamma$-ray strength functions into the TALYS code.
The calculated $^{169}\mathrm{Tm}(\gamma,n)$ rates were adopted in a full $p$-process reaction network for the shock-heated O/Ne-rich layers of a $20 M\odot$ core-collapse supernova model to perform a sensitivity analysis.
The updated rates have a negligible effect on the final $p$-nucleus abundances.

\end{abstract}


\maketitle

\section{Introduction}
The origin of nuclei heavier than iron remains one of the long-standing questions in nuclear astrophysics. Although the $s$- and $r$-processes account for the vast majority of heavy nuclei in the solar system, they do not fully explain the abundances of a group of neutron-deficient isotopes traditionally known as the $p$-nuclei. The classical set comprises 35 isotopes ranging from $^{74}\mathrm{Se}$ to $^{196}\mathrm{Hg}$, most of which are shielded from $r$-process decay chains by stable isobars and bypassed by the $s$-process reaction flow \citep{Burbidge1957,Cameron1957,Arnould2003,Rauscher2013}. A leading mechanism for the production of most $p$-nuclei is the $\gamma$-process, in which pre-existing heavy seed nuclei produced by the $s$- and/or $r$-processes are exposed to an intense thermal photon field. The reaction flow is initiated mainly by $(\gamma,n)$ reactions, which drive material toward the proton-rich side of the nuclear chart; as neutron separation energies increase, competing $(\gamma,p)$ and $(\gamma,\alpha)$ channels redirect the flow toward lower atomic numbers, followed by $\beta^{+}$ decays or electron captures during freeze-out \citep{Arnould2003}. Several astrophysical environments have been investigated as potential $\gamma$-process sites, including the shock-heated O/Ne-rich layers of massive stars during core-collapse supernovae and the outer layers of exploding carbon--oxygen white dwarfs in Type Ia supernovae \citep{Rayet1995,Rauscher2002,Travaglio2011,Travaglio2018,Roberti2023}. Their relative contributions to the solar $p$-nucleus abundances remain uncertain. Complementary mechanisms, including the $\nu p$-process, the $rp$-process, and neutrino-induced reactions, have also been explored to account for selected $p$-nuclei that are difficult to reproduce through the $\gamma$-process alone \citep{Frohlich2006,Rauscher2013}.

\begin{figure*}[ht]
\centering
\includegraphics[width=0.9\linewidth]{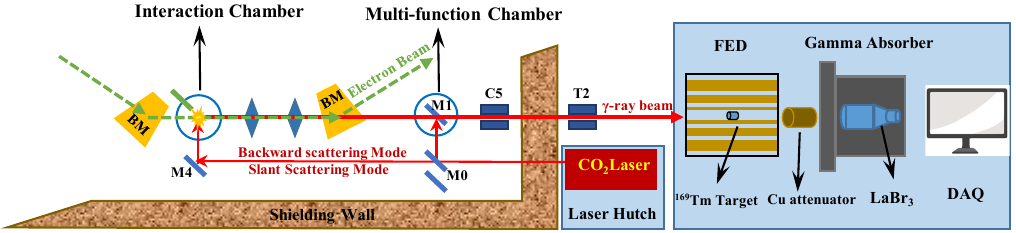}
\caption{Schematic diagram of the photoneutron cross-section measurement setup at the SLEGS beamline.
The quasi-monoenergetic $\gamma$ beam generated by laser Compton scattering irradiates the target at the center of the flat-efficiency neutron detector (FED).
The residual $\gamma$ rays downstream of the target are attenuated and measured with a scintillation detector to determine the incident $\gamma$-ray flux and energy distribution.}
\label{fig:setup}
\end{figure*}

In addition to uncertainties in astrophysical sites, stellar evolution, and seed abundances, nuclear reaction rates constitute a major source of uncertainty in $\gamma$-process nucleosynthesis calculations \citep{Rapp2006,Rauscher2013}. A comprehensive $p$-process network includes neutron-, proton-, and $\alpha$-induced reactions together with their inverse photodisintegration channels. Such networks can involve more than 20,000 reactions linking approximately 2000 stable and unstable nuclei with mass numbers up to $A\approx210$ \citep{Arnould2003}. Experimental information is available for only a small fraction of these reactions because many relevant nuclei are unstable or have very low natural isotopic abundances, while cross sections at astrophysically relevant energies can be extremely small. Consequently, most reaction rates are calculated using statistical Hauser--Feshbach (HF) models. Their predictive reliability depends on the underlying nuclear inputs, notably the $\gamma$-ray strength function ($\gamma$SF), nuclear level density (NLD), which determine the transmission coefficients and competition among reaction channels \citep{Rauscher2013,TALYS,RIPL3}. Measurements of inverse $(p,\gamma)$ and $(\alpha,\gamma)$ reactions in normal and inverse kinematics provide experimental constraints on calculations of $(\gamma,p)$ and $(\gamma,\alpha)$ rates \citep{Fallis2020,PalmisanoKyle2022,Tsantiri2023}. Recent advances have extended proton-capture measurements to radioactive beams, including stored ions \citep{Lotay2021,Dellmann2025}. However, direct photodisintegration data remain limited, particularly near reaction thresholds. Measurements of these cross sections therefore provide valuable benchmarks for testing HF models and constraining the nuclear inputs used to predict stellar reaction rates.

The rare-earth region is of particular interest because several $p$-nuclei with $A\sim150$--165 remain underproduced in standard $\gamma$-process calculations \citep{Arnould2003,Rauscher2013}. Photoneutron measurements in this region provide experimental tests of the statistical models used to predict the relevant reaction rates. For example, Hasper et al. investigated near-threshold photoneutron reactions on Nd, Sm, and Gd isotopes through photoactivation measurements and compared the measured yields with statistical-model predictions \citep{Hasper2008}. Within the rare-earth region, the stable, deformed odd-mass nucleus $^{169}\mathrm{Tm}$ offers an experimentally accessible case for testing the description of photoexcitation and neutron emission from the near-threshold region into the giant dipole resonance (GDR).

Photoneutron cross sections for $^{169}$Tm were previously measured with laser-Compton-scattering (LCS) $\gamma$-ray beams at NewSUBARU, and preliminary results were provided to the IAEA coordinated research project that produced the IAEA-2019 photonuclear data library \cite{IAEA2019}.
The NewSUBARU group has recently analyzed these measurements using updated photon-flux determinations and LCS $\gamma$-beam spectra, together with revised treatments of beam nonmonochromaticity and neutron-multiplicity sorting, yielding revised cross sections \cite{Gheorghe2026,Gheorghe2021NIMA}.
Since the preliminary and updated results originate from the same measurements, an independent determination is particularly valuable for assessing the absolute normalization and energy dependence of the $^{169}$Tm$(\gamma,n)$ excitation function.
A measurement at a different LCS facility, with an independent neutron-detection system, photon-flux determination, and unfolding analysis, therefore provides a complementary data set for examining the consistency of the available cross sections.

In the present work, we measured the $^{169}\mathrm{Tm}(\gamma,n)^{168}\mathrm{Tm}$ cross section using LCS $\gamma$-ray beams at the Shanghai Laser Electron Gamma Source (SLEGS), from just above the neutron emission threshold to 14.61~MeV. This energy range probes the near-threshold response and the lower-energy component of the deformation-split GDR. The measured excitation function is compared with existing data and HF calculations to test the predicted absolute magnitude and energy dependence of the cross section and to examine the sensitivity to the adopted $\gamma$SF and NLD. The experimental constraints are then used to calculate the $^{169}\mathrm{Tm}(\gamma,n)$ reaction rate within the HF framework, with contributions from thermally populated target states included in the evaluation of the stellar rate. Finally, the sensitivity of $p$-nucleus production to this reaction rate is investigated through reaction-network calculations for the O/Ne shell of a $20\,M_{\odot}$ core-collapse supernova model.

Section~\ref{sec:exp} describes the SLEGS experimental setup, including the target, flat-efficiency neutron detector, and $\gamma$-ray beam-monitoring system.
Section~\ref{sec:analysis} presents the cross-section extraction, iterative unfolding procedure, and uncertainty analysis.
Section~\ref{sec:results} discusses the measured excitation function and GDR structure and examines the constraints on the $\gamma$SF within the Hauser--Feshbach framework.
Section~\ref{sec:stellar_rate} presents the ground-state and stellar reaction rates.
Section~\ref{sec:astro} examines the sensitivity of the $p$-process abundances in the $20 M_{\odot}$ core-collapse supernova model to this rate.
Section~\ref{sec:conclusion} summarizes the main conclusions of this work.

\section{Experimental setup}\label{sec:exp}
The SLEGS located at the Shanghai Synchrotron Radiation Facility, produces energy-tunable, quasi-monochromatic $\gamma$-ray beams through laser-Compton scattering.
In its slant-scattering and backscattering configurations, SLEGS provides photon energies ranging from approximately 0.66 to 21.7~MeV, with total photon fluxes of about $10^{5}$--$10^{7}$ photons~s$^{-1}$.
A two-stage collimation system consisting of a coarse collimator (C5, 5~mm aperture) and a fine collimator (T2, 2~mm aperture) defines the beam spot and improves the energy resolution to 6\%--12\% (FWHM) across the energy range of this measurement \cite{Wang2022SLEGS,Hao2025PCSa,Liu2024BGO}.
The beam energy spread can be tuned to a few percent under optimized collimation conditions \cite{Wang2022SLEGS,Hao2025PCSa}.
Recent photo-nuclear measurements at SLEGS have further demonstrated the capability of the facility to discriminate between discrepant data sets among existing experimental data and evaluated nuclear-data libraries \cite{deuteronPRL,HeJJPRL,Rh103SLEGS}.

A schematic of the experimental arrangement is illustrated in Fig.~\ref{fig:setup}.
The shaped $\gamma$-ray beam was guided to impinge precisely on a $^{169}$Tm target located at the center of a $4\pi$ flat-efficiency neutron detector (FED) array.
Neutrons emitted via the $(\gamma,n)$ reaction were registered with high efficiency, while the residual $\gamma$-ray flux and energy profile were monitored downstream using a large-volume LaBr$_3$ total-absorption spectrometer installed in the beam dump.

\begin{figure}[htbp]
\centering
\includegraphics[width=\linewidth]{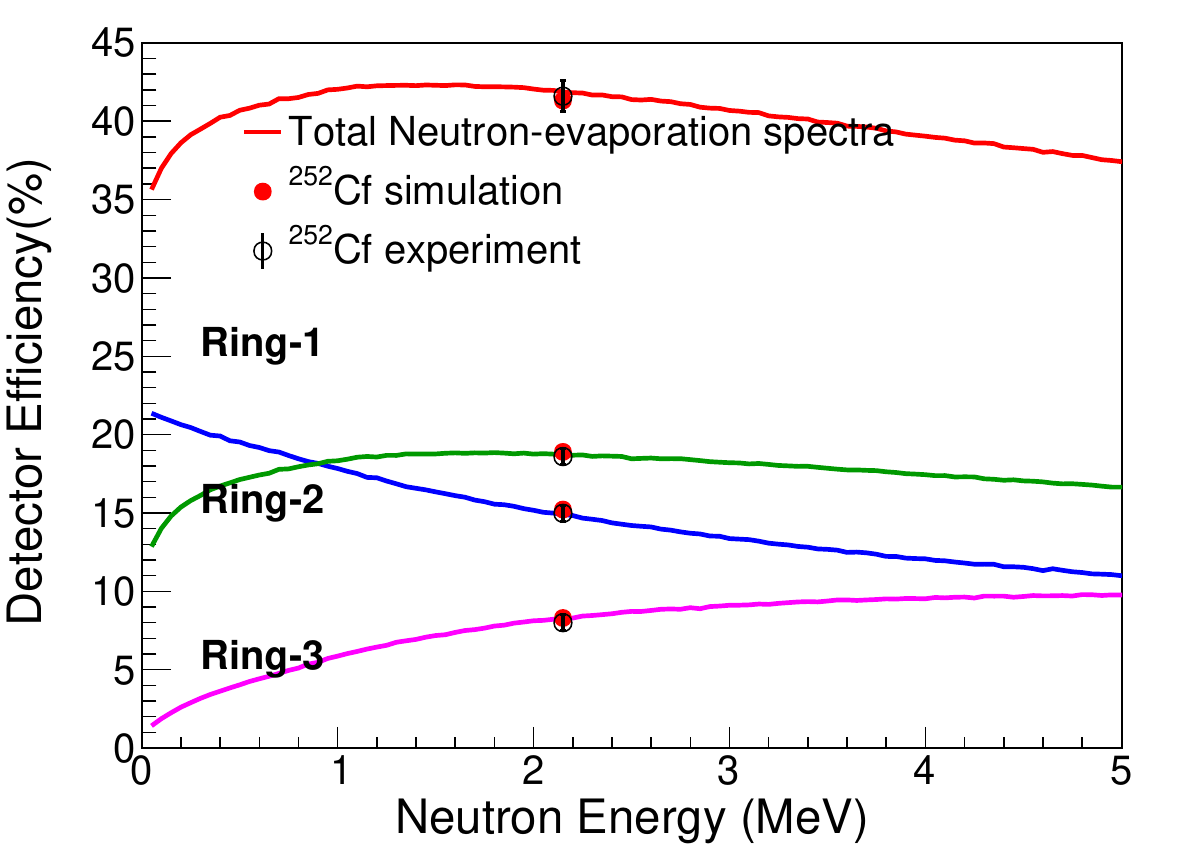}
\caption{Simulated neutron detection efficiency of the FED array. The efficiencies of the individual detector rings (R1, R2, and R3) as well as the total efficiency of the detector system are shown.}
\label{fig:efficiency}
\end{figure}

\subsection{Target preparation and assembly}
A high-purity metallic $^{169}$Tm target was used for the photoneutron measurements.
The target material had a chemical purity better than 99.99\%, with trace-element impurities below 100~ppm.
The target was fabricated as a circular disk with a diameter of 10~mm and a thickness of 5.0~mm, corresponding to an areal density of $1.66\times10^{22}$~nuclei/cm$^2$.
The mass density of thulium was taken as 9.32~g/cm$^3$.
The target mass was determined with a calibrated analytical balance with a resolution of 0.01~mg, and the diameter and thickness were measured with a micrometer at several positions across the disk; the resulting relative uncertainty of the areal density is below 0.1\% (see Sec.~\ref{sec:uncertainty}).

The target was mounted at the geometric center of the $4\pi$ flat-efficiency neutron detector (FED) array and precisely aligned with the LCS $\gamma$-ray beam using a miniaturized X-ray imaging system to ensure full overlap between the beam and the target.
The target holder was made of low-activation polyethylene, with a 10-mm-diameter aperture to minimize neutron background from the holder material.

\subsection{Flat-efficiency neutron detector}\label{sec:fed}
Photoneutrons emitted from the $^{169}$Tm target were detected using the SLEGS $4\pi$ flat-efficiency $^3$He counter array (FED).
The FED is specifically designed for photonuclear experiments and provides nearly constant detection efficiency over a wide neutron energy range \cite{FEDDesign}.
The detector system consists of 26 cylindrical $^3$He proportional counters embedded in a compact polyethylene moderator assembly of dimensions $450\times450\times550$~mm$^3$.
The counters are arranged in three concentric rings at radii of 6.5~cm (inner ring, 6 counters), 11.0~cm (middle ring, 8 counters), and 17.5~cm (outer ring, 12 counters) relative to the beam axis.
All counters are 500~mm in length and filled with $^3$He gas at 2~atm pressure.

The FED is surrounded by cadmium-loaded polyethylene shielding to suppress thermal and epithermal background neutrons.
The efficiency was calibrated using a calibrated $^{252}$Cf radioactive source and validated using the ring-ratio technique and Geant4 simulations \cite{Hao2021}.
Figure~\ref{fig:efficiency} shows the calculated detection efficiencies for the individual detector rings as well as the total efficiency of the entire detector system.
As shown in Fig.~\ref{fig:efficiency}, the efficiencies of the individual detector rings vary with neutron energy because of their different distances from the target and the moderation process in the polyethylene.
The summed efficiency of the full detector array, however, varies only mildly, from about 35\% for thermal neutrons to about 42\% for 1~MeV neutrons, and is nearly constant over the neutron-energy range relevant to this measurement.
This flat response reduces the sensitivity of the extracted $(\gamma,n)$ cross section to the outgoing-neutron energy distribution, which is not known a priori in the GDR region \cite{Hao2021}.

The efficiency value adopted in the cross-section extraction is the spectrum-averaged efficiency,
\begin{equation}
\epsilon_n = \frac{\int \phi_n(E_n)\,\epsilon(E_n)\,dE_n}{\int \phi_n(E_n)\,dE_n},
\label{eq:eff_avg}
\end{equation}
where $\epsilon(E_n)$ is the simulated energy-dependent total efficiency and $\phi_n(E_n)$ is the expected $(\gamma,n)$ neutron spectrum, taken from the TALYS-2.0 Hauser--Feshbach calculation with the experimentally constrained $\gamma$SF.
Because the neutrons from the $(\gamma,n)$ reaction in the measured energy region are emitted predominantly with energies below $\sim$1.5~MeV, the averaging weights the efficiency toward the flat portion of the response, yielding $\epsilon_n \approx 40\%$.
To quantify the residual model dependence, the averaging was repeated with Maxwellian spectra of different temperatures and with several NLD and $\gamma$SF prescriptions; the resulting spread of $\epsilon_n$ is within 3\%.
The adopted 3.02\% uncertainty is the quadratic sum of the $^{252}$Cf calibration uncertainty, the ring-ratio validation spread, and this spectrum-averaging spread, as detailed in previous SLEGS analyses \cite{Hao2021,Hao2025PCSa}.
The ring-resolved efficiencies vary more strongly with energy, by about $\pm$8\% between their extremes.
That variation does not carry over to the total efficiency, because the spectrum average is taken over the neutron energies actually emitted in the present measurement.

\subsection{$\gamma$-ray beam monitor and spectrometer}

Accurate knowledge of the incident $\gamma$-ray energy distribution is essential for extracting photonuclear cross sections from experimental data.
Because the LCS $\gamma$-ray beam is not strictly monochromatic, the energy distribution of the beam must be measured and unfolded in order to determine the effective photon energy for each experimental point.
In the present experiment, the $\gamma$-ray spectrum was measured using a LaBr$_3$(Ce) scintillation detector placed downstream of the target position.
LaBr$_3$ detectors provide excellent timing performance and relatively good energy resolution in the MeV region, making them well suited for monitoring high-energy $\gamma$ rays \cite{LaBr3Review}.
The measured pulse-height spectra recorded by the LaBr$_3$ detector include contributions from the detector response, such as Compton scattering and pair production processes.
Before reaching the LaBr$_3$ detector, the $\gamma$ beam passed through a calibrated copper attenuator to reduce the instantaneous count rate and avoid pileup.
Therefore, the measured spectra must be corrected for detector response effects to obtain the true $\gamma$-ray energy distribution \cite{Liu2024BGO}.

\begin{figure}[htbp]
\centering
\includegraphics[width=\linewidth]{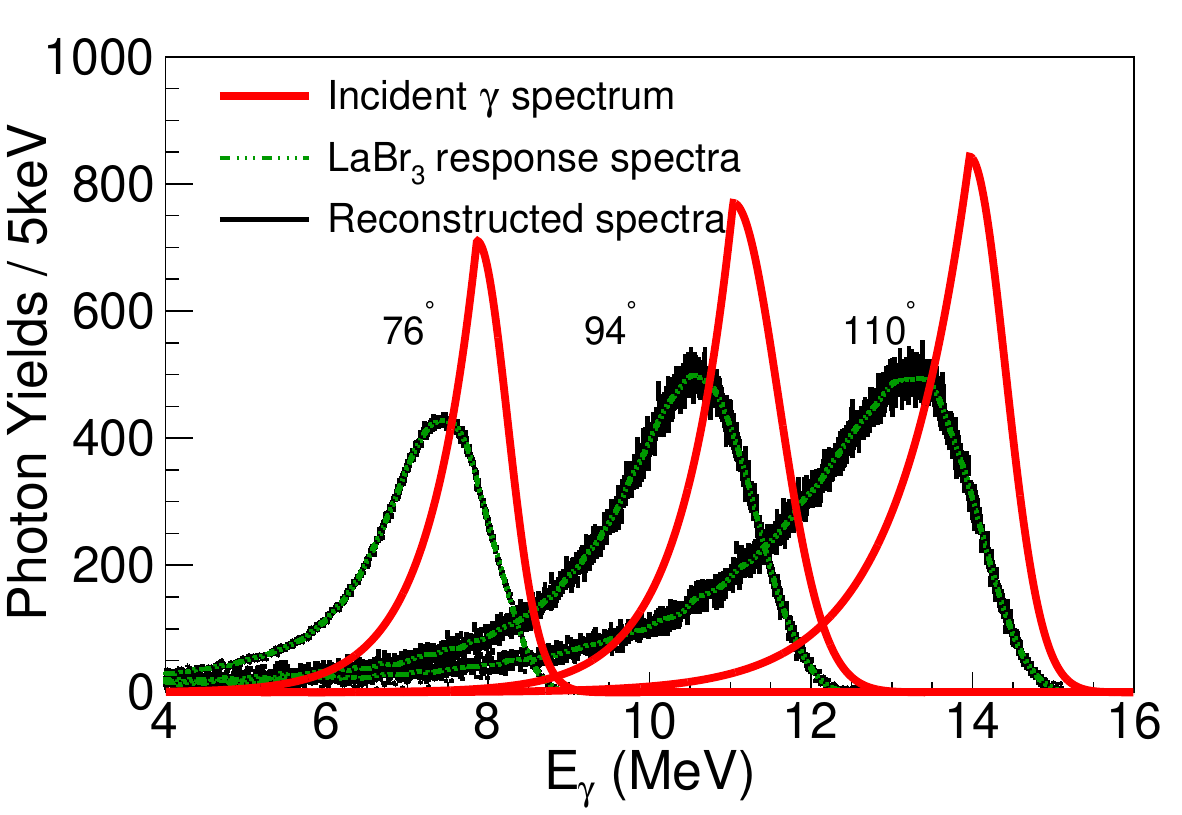}
\caption{Typical $\gamma$-ray spectra measured by the LaBr$_3$ detector at different observation angles. Green dotted lines: measured LaBr$_3$ response spectra; black lines: response-folded spectra of the reconstructed incident
spectra, which reproduce the measured ones; red lines: reconstructed incident $\gamma$-ray spectra.}
\label{fig:gamma_spectrum}
\end{figure}

Figure~\ref{fig:gamma_spectrum} shows examples of the measured $\gamma$-ray spectra obtained at different observation angles.
The green dotted lines represent the raw spectra measured by the LaBr$_3$ detector.
After applying the response correction and unfolding procedure, the corresponding folded spectra and reconstructed incident $\gamma$ spectra were obtained.
The reconstructed spectra represent the actual $\gamma$-ray energy distributions incident on the target.
The reconstructed $\gamma$ spectra were subsequently used in the unfolding procedure to determine the energy-dependent photoneutron cross sections.
The uncertainty of the $\gamma$-ray flux determination has two origins: the calibration of the LaBr$_3$ detector and the copper attenuator, and the response-unfolding procedure that converts the measured pulse-height spectra into incident photon spectra.
The latter is based on an iterative response-removal technique of the type established in Ref.~\cite{OsloMethod}, whose convergence and stability for scintillator spectra have been demonstrated there; the former is quantified from the calibration-source measurements and the attenuator characterization described in Ref.~\cite{Liu2024BGO}.

\subsection{Data acquisition system}

Signals from the $^{3}$He counters were amplified and digitized with Mesytec MDPP-16 modules, then recorded in list mode by the MVME system.
Pulse amplitude, channel, and timing were stored for each event and converted to ROOT format offline.
The LaBr$_3$ signals were digitized with a CAEN DT5730B unit running CoMPASS.

The synchronization between neutron and $\gamma$ detection systems is achieved using the timing signals associated with the laser pulses that generate the $\gamma$ rays.
Because the laser operates in a pulsed mode, the resulting photoneutron events exhibit a characteristic time structure that allows photon-induced neutrons to be distinguished from background radiation \cite{Hao2021}.

\section{Data analysis}\label{sec:analysis}
\subsection{Cross-section extraction}

The cross-section extraction and iterative unfolding procedure follow the method established in our recent SLEGS photoneutron studies \cite{deuteronPRL,Cu63SLEGS}; here we summarize the key steps and highlight the aspects specific to the present $^{169}$Tm measurement.

The folded photoneutron cross section for an incoming $\gamma$-ray beam with maximum energy $E_{\rm max}$ is obtained from the measured neutron yield via
\begin{equation}
\label{eq:xs}
\int_{S_n}^{E_{\rm max}} n_\gamma(E_\gamma)\,\sigma(E_\gamma)\,dE_\gamma
= \frac{N_n}{N_\gamma N_t \zeta \epsilon_n},
\end{equation}
where $S_n = 8.03$~MeV is the neutron separation energy of $^{169}$Tm,
$n_\gamma(E_\gamma)$ is the normalized LCS photon energy distribution,
$N_n$ is the background-subtracted neutron count, $N_\gamma$ is the
incident photon flux, $N_t$ is the areal target density,
$\epsilon_n$ is the FED detection efficiency, and $\zeta$ is the
thick-target self-attenuation factor.
Because all energies lie below the $(\gamma,2n)$ threshold
($S_{2n} = 14.88$~MeV), no neutron-multiplicity correction is needed.

For the present experiment, the incident photon flux, continuously monitored with the LaBr$_3$ spectrometer and the calibrated copper attenuator, was of the order of $10^{5}$ photons~s$^{-1}$, within the $10^{5}$--$10^{7}$ photons~s$^{-1}$ range quoted for the slant-scattering mode of SLEGS.
The flux was stable during each individual run; residual long-term variations were tracked run by run by the LaBr$_3$ monitor and are included in the normalization of each data point.
The FED efficiency entering Eq.~(\ref{eq:xs}) is the spectrum-averaged value $\epsilon_n \approx 40\%$ defined in Eq.~(\ref{eq:eff_avg}).

\subsection{Iterative unfolding procedure}

The finite energy spread of the LCS beam---arising from electron-beam emittance and collimator acceptance---introduces a low-energy tail into each nominal photon spectrum.
Consequently, each measured folded cross section $\sigma_f$ is a weighted integral of the true monoenergetic cross section $\sigma$ over the reconstructed photon energy distribution $n_\gamma(E_\gamma)$ [Eq.~(\ref{eq:xs})].
Recovering $\sigma$ from the set of $\sigma_f$ values therefore requires inverting the linear relation
\begin{equation}
\boldsymbol{\sigma}_f = \mathbf{D}\,\boldsymbol{\sigma},
\end{equation}
where the response matrix $\mathbf{D}$ has elements $D_{ij}$ equal to the fractional photon flux of the $i$-th LCS setting that falls in the $j$-th energy bin between $S_n$ and $E_{\rm max}$.
Explicitly,
\begin{equation}
\begin{pmatrix}
\sigma_{f1} \\
\sigma_{f2} \\
\vdots \\
\sigma_{fN}
\end{pmatrix}
=
\begin{pmatrix}
D_{11} & D_{12} & \cdots & D_{1M} \\
D_{21} & D_{22} & \cdots & D_{2M} \\
\vdots & \vdots & \ddots & \vdots \\
D_{N1} & D_{N2} & \cdots & D_{NM}
\end{pmatrix}
\begin{pmatrix}
\sigma_1 \\
\sigma_2 \\
\vdots \\
\sigma_M
\end{pmatrix},
\end{equation}
with $N$ LCS beam settings and $M$ energy bins.

\begin{figure}[htbp]
\centering
\includegraphics[width=\linewidth]{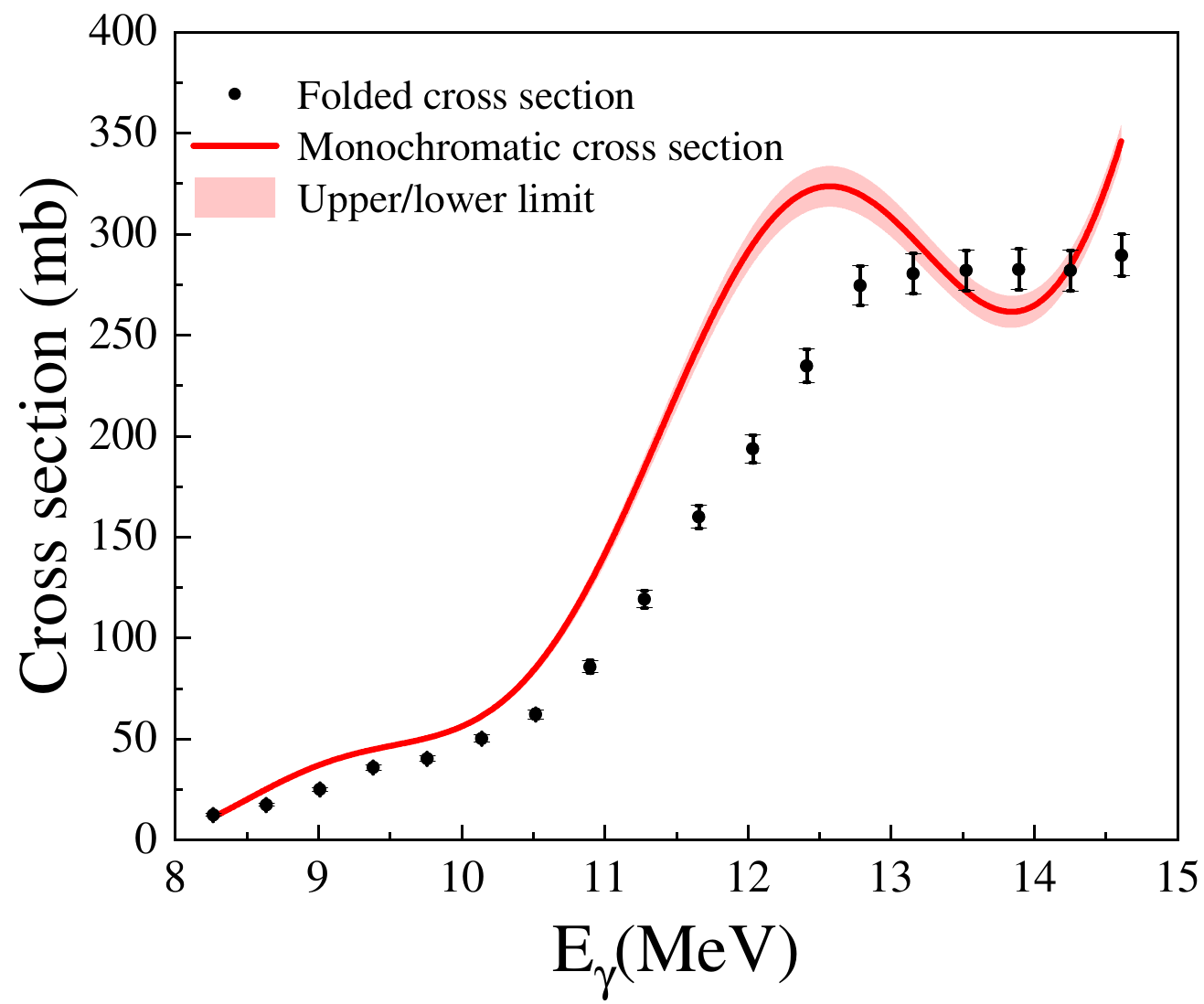}
\caption{Comparison between quasi-monoenergetic cross sections and unfolded monoenergetic cross sections obtained using the iterative unfolding method. The points denote the quasi-monoenergetic cross sections, and the solid line with the shaded band represents the unfolded monoenergetic cross section together with its total uncertainty.}
\label{fig:unfolding}
\end{figure}

Because the system is generally underdetermined ($N < M$), we solve it with the iterative scheme of Ref.~\cite{Guttormsen1996} as applied to LCS data in Ref.~\cite{Renstrom2018}.
Starting from a flat initial trial $\boldsymbol{\sigma}^0$, each cycle computes the predicted folded cross section $\boldsymbol{\sigma}_f^i = \mathbf{D}\boldsymbol{\sigma}^i$ and updates the trial via
\begin{equation}
\boldsymbol{\sigma}^{i+1} = \boldsymbol{\sigma}^i
+ \bigl(\boldsymbol{\sigma}_{f,\rm exp} - \boldsymbol{\sigma}_f^i\bigr),
\end{equation}
with spline interpolation used to reconcile the differing bin structures of $\boldsymbol{\sigma}^i$ and $\boldsymbol{\sigma}_f^i$.
Iterations continue until the reduced $\chi^2$ of $\boldsymbol{\sigma}_f^{i+1}$ relative to the data vector $\boldsymbol{\sigma}_{f,\rm exp}$ reaches unity, signalling that the residuals are consistent with the measurement uncertainties.
The resulting unfolded cross sections for $^{169}$Tm$(\gamma,n)^{168}$Tm, together with their propagated total uncertainties, are shown in Fig.~\ref{fig:unfolding}.

For $^{169}$Tm, the unfolded cross sections are systematically higher than the folded values near threshold, where the low-energy tail of the LCS beam contributes a larger fractional correction.
The shaded band in Fig.~\ref{fig:unfolding} represents the total propagated uncertainty, dominated at low energies by counting statistics and at higher energies by the systematic uncertainty on the FED detection efficiency.

\subsection{Uncertainty analysis}\label{sec:uncertainty}
The total uncertainty of the $^{169}\mathrm{Tm}(\gamma,n)^{168}\mathrm{Tm}$ cross section is composed of three independent components: statistical, algorithmic, and systematic.
The statistical uncertainty originates from the counting statistics of photoneutrons and background events.
For each ring of the FED detector, the statistical error is given by
\begin{equation}
\Delta N_{R_k}^{\text{stat}} = \sqrt{N_{R_k} + N_{R_k}^{\text{bg}}},
\end{equation}
where $N_{R_k}$ and $N_{R_k}^{\text{bg}}$ are the net and background neutron counts of the $k$-th ring.
The statistical uncertainty of the cross section is typically below 3\% in the
GDR peak region.
The algorithmic uncertainty has two components: (1) the variation of the
background-subtraction window yields a 2.20\% uncertainty on
$N_n$~\cite{Hao2025PCSa,deuteronPRL}; (2) the direct unfolding of the LaBr$_3$
response introduces a 1.00\% uncertainty on the incident photon flux
$N_\gamma$~\cite{Liu2024BGO}. The total algorithmic uncertainty is
$\sim$2.20\%.
The systematic contributions are:
\begin{itemize}
\item FED neutron detection efficiency: 3.02\%, which is the quadratic sum of
the $^{252}$Cf calibration uncertainty, the ring-ratio validation spread, and
the spectrum-averaging spread (Sec.~\ref{sec:fed});
\item photon flux determination (LaBr$_3$ calibration and attenuation): 1.00\%;
\item target thickness, areal density, and self-attenuation: $<0.10\%$;
\item copper attenuator impurities and alignment: $<1.00\%$.
\end{itemize}
The areal-density uncertainty follows from the mass and dimensional
measurements described in Sec.~\ref{sec:exp}. Propagating those uncertainties,
and varying the target thickness in the Geant4 self-attenuation calculation,
gives a contribution below 0.1\%.

These percentages refer to the folded cross sections before unfolding.
The uncertainties listed in Table~\ref{tab:cross_section} are the final values
after the iterative unfolding: each component was propagated through the
unfolding procedure (Sec.~\ref{sec:analysis}), so that the values in the table,
and their ratios to the unfolded cross sections, vary with energy and differ
from the percentages quoted above. The final total uncertainty is obtained as
\begin{equation}
\Delta\sigma_{\text{tot}} =
\sqrt{(\Delta\sigma_{\text{stat}})^2+
(\Delta\sigma_{\text{algo}})^2+
(\Delta\sigma_{\text{syst}})^2},
\label{eq:total_unc}
\end{equation}
and ranges from 2.3\% to 6.9\% of the unfolded cross section over the measured
energy range, as listed in Table~\ref{tab:cross_section}. The relative
uncertainty is largest for the lowest-energy point, where counting statistics
and the threshold correction matter most, and it is about 3\%--4\% through
most of the giant dipole resonance region.

The energy axis of the unfolded cross section carries an uncertainty.
The photon energy scale is fixed by the accelerator electron energy and the laser wavelength, and the LaBr$_3$ energy calibration, anchored to standard calibration lines, contributes an uncertainty of a few tens of keV.
The dominant energy-scale effect, however, is the finite width (6\%--12\% FWHM) of the LCS beam itself, which is not an uncertainty on a pointwise energy but is fully accounted for by the response matrix in the unfolding procedure.
The stability of the iterative solution against variations of the response matrix within the calibration uncertainty was verified, and the resulting effect on the cross-section scale is included in the algorithmic uncertainty.

\begin{figure}[h]
\centering
\includegraphics[width=\linewidth]{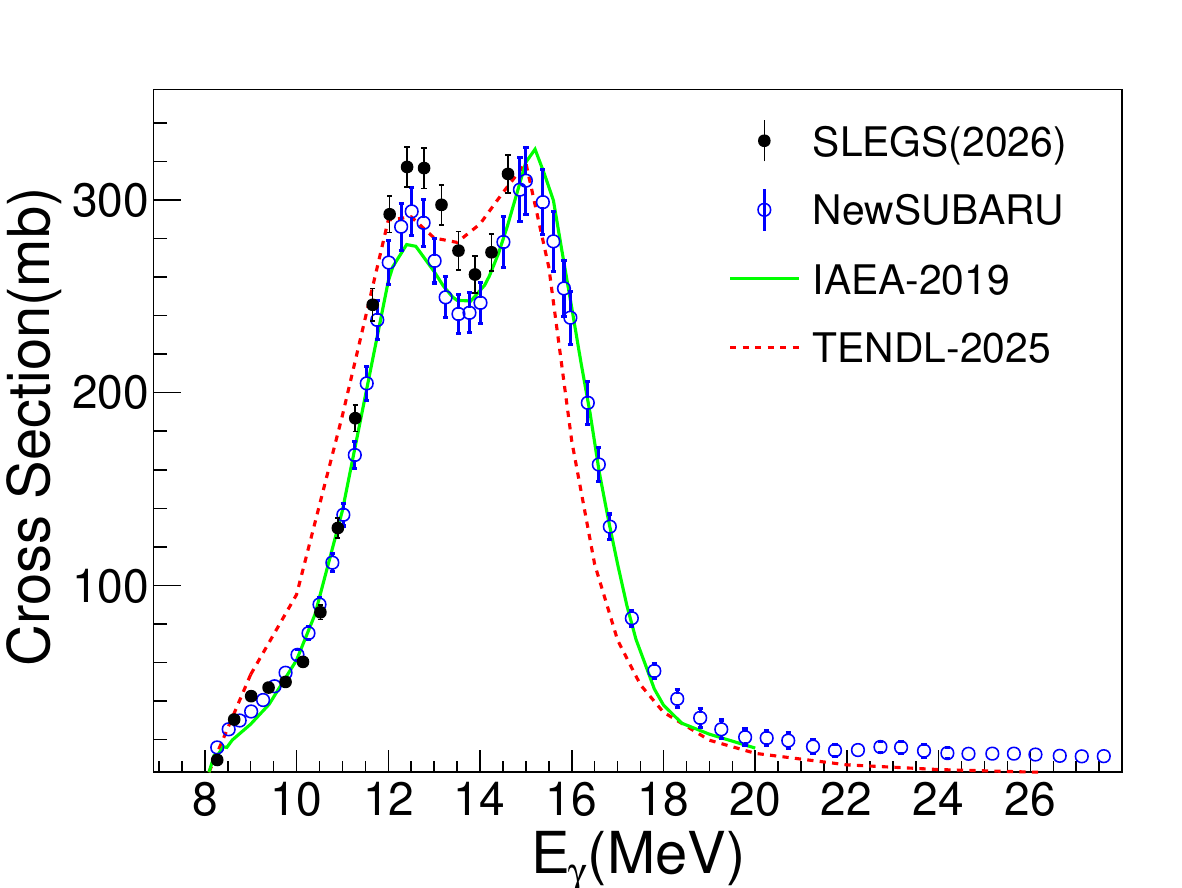}
\caption{Present $^{169}$Tm$(\gamma,n)$ cross section compared with the NewSUBARU $\sigma(\gamma,1nX)$ data, the IAEA-2019 evaluation, and TENDL-2025.}
\label{fig:library}
\end{figure}

\section{Results and discussion}\label{sec:results}

\subsection{Photoneutron cross sections}
The monoenergetic cross sections of the $^{169}$Tm$(\gamma,n)^{168}$Tm reaction were obtained using the iterative unfolding procedure described in Sec.~\ref{sec:analysis}.
The extracted cross sections and their statistical, algorithmic, systematic, and total uncertainties are listed in Table~\ref{tab:cross_section}, and the resulting excitation function is shown in Fig.~\ref{fig:library}.
The cross section increases rapidly above the neutron threshold and reaches a broad maximum in the giant dipole resonance (GDR) region.
This behavior is consistent with the dipole response expected for a deformed rare-earth nucleus, in which the strength is distributed over a relatively broad energy interval \cite{Berman1975,BohrMottelson,Goriely2004}.

\begin{table}[b]
\centering
\caption{Measured $^{169}$Tm$(\gamma,n)^{168}$Tm cross sections obtained by unfolding (monoenergetic cross sections). Stat., Algo., and Sys.\ denote the statistical, algorithmic, and systematic uncertainties after unfolding, respectively.}
\label{tab:cross_section}
\begin{tabular}{cccccc}
\hline\hline
Energy & Cross section & Stat. & Algo. & Sys. & Total \\
(MeV) & (mb) & (mb) & (mb) & (mb) & (mb) \\
\hline
8.27 & 9.48 & 0.63 & 0.09 & 0.11 & 0.65 \\
8.64 & 30.45 & 0.45 & 0.29 & 0.50 & 0.73 \\
9.01 & 42.57 & 0.32 & 0.46 & 0.83 & 1.00 \\
9.39 & 47.03 & 0.25 & 0.64 & 1.18 & 1.37 \\
9.76 & 49.96 & 0.23 & 0.87 & 1.65 & 1.88 \\
10.14 & 60.35 & 0.26 & 1.22 & 2.37 & 2.68 \\
10.52 & 86.08 & 0.33 & 1.73 & 3.39 & 3.82 \\
10.90 & 129.80 & 0.42 & 2.37 & 4.69 & 5.27 \\
11.28 & 186.75 & 0.52 & 3.08 & 6.11 & 6.86 \\
11.66 & 245.57 & 0.60 & 3.76 & 7.45 & 8.37 \\
12.03 & 292.53 & 0.64 & 4.31 & 8.50 & 9.55 \\
12.41 & 317.08 & 0.65 & 4.66 & 9.12 & 10.26 \\
12.78 & 316.49 & 0.62 & 4.79 & 9.30 & 10.48 \\
13.16 & 297.42 & 0.55 & 4.75 & 9.15 & 10.32 \\
13.53 & 273.68 & 0.48 & 4.62 & 8.85 & 9.99 \\
13.89 & 261.29 & 0.41 & 4.49 & 8.58 & 9.69 \\
14.25 & 272.81 & 0.38 & 4.45 & 8.50 & 9.60 \\
14.61 & 313.45 & 0.39 & 4.51 & 8.66 & 9.77 \\
\hline\hline
\end{tabular}
\end{table}

The present results are compared with available experimental and evaluated data in Fig.~\ref{fig:library}, including the updated NewSUBARU LCS data~\cite{Gheorghe2026}, the IAEA-2019 evaluation~\cite{IAEA2019}, and the TENDL-2025 evaluation~\cite{TENDL2025}. Below 12~MeV, the present results agree well with the NewSUBARU data and the IAEA-2019 evaluation. Around the lower-energy maximum of the excitation function, however, the present cross sections are higher than both. The IAEA-2019 evaluation was based on preliminary NewSUBARU results preceding the recent analysis and thus shares an experimental basis with the updated NewSUBARU data \citep{IAEA2019,Gheorghe2026}. Differences in photon-flux determination, neutron-detection efficiency, energy calibration, target-thickness corrections, and unfolding procedures may contribute to the remaining discrepancy \citep{Gheorghe2026Tb159,Gheorghe2021NIMA}. The independent SLEGS measurement therefore provides a complementary benchmark for assessing the absolute normalization and energy dependence of the excitation function. TENDL-2025 does not satisfactorily reproduce the measured shape: it overestimates the low-energy rising slope, underestimates the lower-energy maximum, and predicts a less pronounced minimum between the GDR components.

\subsection{Lorentzian fit to the GDR}
\begin{figure}[htbp]
    \centering
    \includegraphics[width=\linewidth]{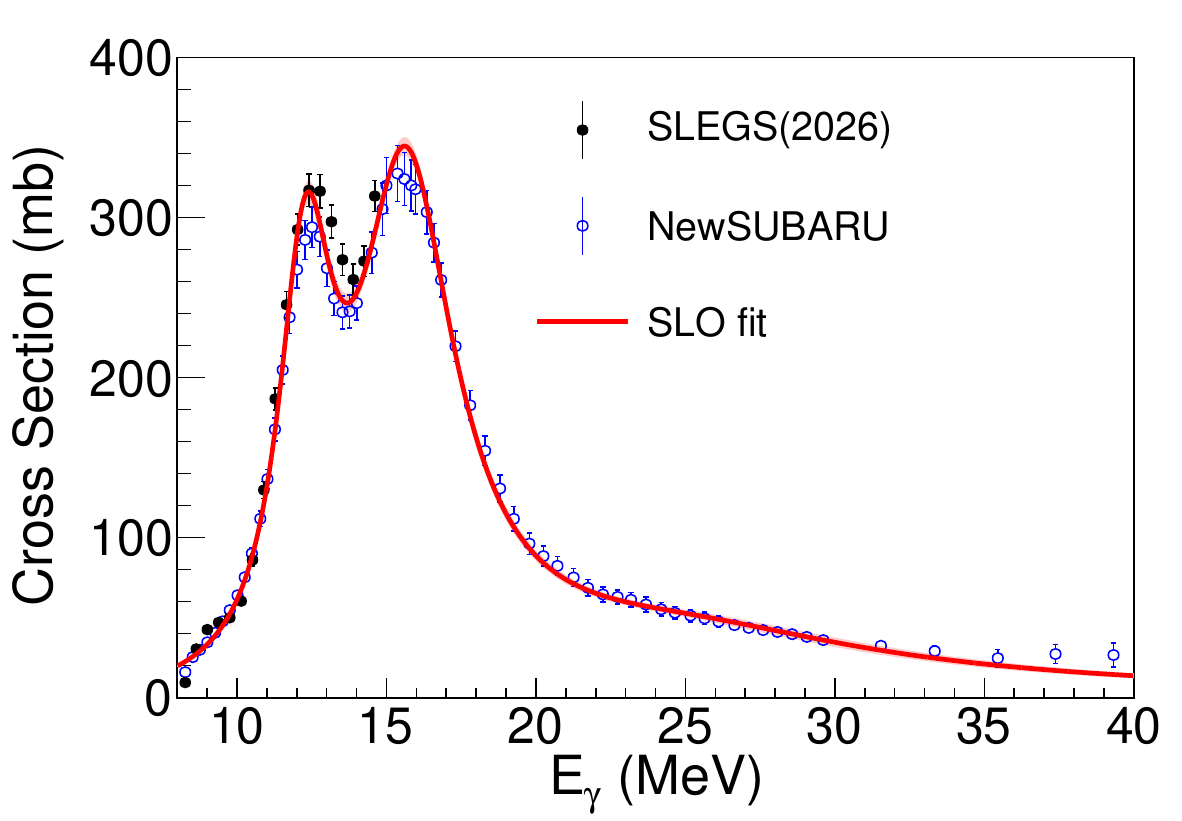}
    \caption{Total photoneutron cross section $\sigma(\gamma,S_n)$ of $^{169}$Tm measured at SLEGS (present data, $(\gamma,n)$ below $S_{2n}$) and NewSUBARU. The red solid curve is the three-component SLO (Brink--Axel) fit to both data sets jointly, following Ref.~\cite{Berman1975}. The shaded band represents the $1\sigma$ uncertainty of the joint fit, propagated from the covariance matrix of the fitted parameters.}

    \label{fig:gdrfit}
\end{figure}

The photoabsorption cross section is commonly described by a sum of standard Lorentzian (SLO) functions~\cite{Berman1975,RIPL3}. This is the form associated with the Brink--Axel hypothesis~\cite{Brink1955,Axel1962}; in TALYS it is called the Brink--Axel option, in which a standard Lorentzian (SLO)
describes the GDR shape~\cite{TALYS}. The total photoneutron cross section $\sigma(\gamma,S_n)$ was fitted with
\begin{equation}
\sigma(E)=\sum_{i=1}^{3}\frac{\sigma_{mi}E^{2}\Gamma_i^{2}}
{E^{2}\Gamma_i^{2}+(E^{2}-E_{mi}^{2})^{2}},
\label{eq:slo}
\end{equation}
using all present data points (8.27--14.61~MeV, where $\sigma(\gamma,n)=\sigma(\gamma,S_n)$) together with the NewSUBARU $\sigma(\gamma,S_n)$ data (8.26--39.3~MeV), each weighted by its quoted uncertainty.
The resulting joint (SLEGS + NewSUBARU) fit is shown by the red solid curve in Fig.~\ref{fig:gdrfit}, and the fitted parameters are listed in Table~\ref{tab:fitpar}.

To examine how the present data constrain the GDR, the same three-component SLO form was also fitted to the SLEGS data alone; the results are included in Table~\ref{tab:fitpar}.
 The SLEGS data extend only to 14.61~MeV: they fully cover the first GDR component but only the low-energy flank of the second one ($E_{m2}\simeq 15.7$~MeV), whose centroid and width therefore cannot be determined from the SLEGS data alone. In the SLEGS-only fit, $E_{2}$ and $\Gamma_{2}$ were fixed to the values of Ref.~\cite{Gheorghe2026} ($15.72$ and $4.25$~MeV) and the third component to the joint-fit values, leaving free the three parameters of the first component and the peak cross section $\sigma_{m2}$ of the second one, the latter being constrained by the flank data.

\begin{table}[htbp]
\centering
\caption{Parameters of the three-component SLO (Brink--Axel) fit to the total photoneutron cross section $\sigma(\gamma,S_n)$ of $^{169}$Tm: joint fit to all data, fit to the SLEGS data alone with $E_{2}$ and $\Gamma_{2}$ fixed to Ref.~\cite{Gheorghe2026} and the third component fixed to the joint-fit values. $\sigma_{mi}$, $E_{mi}$, and $\Gamma_i$ are the peak cross section, resonance energy, and width of the $i$-th Lorentzian. The Ref.~\cite{Gheorghe2026} values are quoted without uncertainties, as in the original.}
\label{tab:fitpar}
\begin{tabular}{ccccc}
\hline\hline
Fit & $i$ & $\sigma_{mi}$ (mb) & $E_{mi}$ (MeV) & $\Gamma_{i}$ (MeV) \\
\midrule
SLEGS & 1 & $247.86 \pm 5.14$ & $12.30 \pm 0.03$ & $2.21 \pm 0.06$ \\
 +NewSUBARU& 2 & $306.95 \pm 5.56$ & $15.68 \pm 0.04$ & $3.90 \pm 0.12$ \\
 & 3 & $30.46 \pm 1.40$ & $25.38 \pm 0.54$ & $15.23 \pm 1.34$ \\
\midrule
SLEGS only & 1 & $270.71 \pm 6.57$ & $12.33 \pm 0.03$ & $2.20 \pm 0.06$ \\
 & 2 & $305.45 \pm 12.60$ &  - & - \\
& 3 & - & - & - \\

\midrule
NewSUBARU~\cite{Gheorghe2026} & 1 & $222.0$ & $12.28$ & $2.38$ \\
 & 2 & $292.0$ & $15.72$ & $4.25$ \\
 & 3 & $20.82$ & $25.18$ & $12.00$ \\
\hline\hline
\end{tabular}
\end{table}

The shaded band in Fig.~\ref{fig:gdrfit} represents the $1\sigma$ uncertainty of the joint fit, propagated from the covariance matrix of the nine fitted parameters. The published three-component SLO fit of Ref.~\cite{Gheorghe2026} to the NewSUBARU data is not drawn; its parameters are included in Table~\ref{tab:fitpar} for comparison.

The joint fit is not an average of the two single-data-set results. Because the two Lorentzians overlap strongly at 12--15~MeV, where the SLEGS cross sections lie systematically above the NewSUBARU ones, the joint fit gives a smaller $\sigma_{m1}$ than the SLEGS-only fit, while $\sigma_{m2}$ remains consistent with the SLEGS-only value within the uncertainties.
The large value $\chi^{2}/\mathrm{ndf}\simeq8.7$ reflects the tension between the two data sets; the parameter uncertainties in Table~\ref{tab:fitpar} are therefore lower limits.

The three Lorentzian components are not constrained equally well by the data.
The first two components, which describe the split GDR, are determined by the present SLEGS data and by the NewSUBARU points in the 8--20~MeV region, where the uncertainties are at the level of a few percent.
In contrast, the third component is constrained solely by the NewSUBARU data above 20~MeV~\cite{IAEA2019,Gheorghe2026}, which are sparse and have large uncertainties.
In particular, only five points are available above 30~MeV (31.5--39.3~MeV), with relative uncertainties ranging from about 9\% to 41\%, the largest being that of the highest-energy point.
In this region the cross section first decreases smoothly and then rises again between about 35 and 39~MeV; part of this rise is related to the growth of the $(\gamma,4nX)$ channel, whose cross section increases from about 0.3 to 9~mb between 31.5 and 39.3~MeV.
The broad third component ($E_{m3}\simeq 25$~MeV, $\Gamma_3\simeq 15$~MeV) acts as an effective term absorbing this high-energy behavior, including the high-energy structure reported in Ref.~\cite{Gheorghe2026}.
Its parameters carry large uncertainties and are strongly correlated, so we regard it as a phenomenological parameterization and do not attribute physical significance to its individual values.
Its contribution in the GDR region amounts to only a few mb, hence it has no appreciable effect on the description of the energy range relevant for the $\gamma$SF and the stellar rate ($E_\gamma\lesssim 15$~MeV).
The joint-fit parameters of Table~\ref{tab:fitpar} are adopted for all subsequent calculations in this work. They define the constrained E1 $\gamma$SF of Sec.~\ref{sec:gsf} and provide the $\gamma$SF input to the TALYS-2.0 calculations of the $^{168}$Tm$(n,\gamma)$ cross section and of the stellar rates in Sec.~\ref{sec:stellar_rate}. The SLEGS-only and Ref.~\cite{Gheorghe2026} parameter sets are used only for comparison.

\subsection{Constraint on the $\gamma$-ray strength function}
\label{sec:gsf}

The $\gamma$SF characterizes the average electromagnetic transition strength and determines the corresponding photon transmission coefficients.
It therefore plays a central role in Hauser--Feshbach (HF) calculations of photonuclear and radiative-capture cross sections \cite{Hauser1952,TALYS}.
Various phenomenological and microscopic prescriptions have been developed for the $\gamma$SF, and experimental data are essential for testing these models and constraining the strength function.

For dipole radiation, the downward strength function associated with electromagnetic de-excitation can be written as \cite{Kopecky1990}
\begin{equation}
\overleftarrow{f}_{X1}(E_{\gamma}) =
E_{\gamma}^{-3}
\frac{\langle \Gamma_{X1}(E_{\gamma}) \rangle}
{D_{\lambda}},
\end{equation}
where $X$ denotes electric ($E$) or magnetic ($M$) dipole radiation, $\langle \Gamma_{X1}(E_{\gamma}) \rangle$ is the average radiative width, and $D_{\lambda}$ is the average level spacing of the corresponding compound states.

For photoabsorption, the upward dipole strength function is related to the average E1 or M1 photoabsorption cross section through
\begin{equation}
\overrightarrow{f}_{X1}(E_{\gamma}) =
\frac{1}{3(\pi\hbar c)^2}
\frac{\langle \sigma_{X1}^{\mathrm{abs}}(E_{\gamma}) \rangle}
{E_{\gamma}} .
\end{equation}

Within the generalized Brink--Axel approximation, the upward and downward strength functions are assumed to provide approximately the same average electromagnetic strength at a given transition energy \cite{Brink1955,Axel1962}.
In the energy region above the neutron separation energy where neutron emission dominates the photoabsorption decay, the E1 strength can therefore be constrained from the measured photoneutron cross section according to
\begin{equation}
f_{E1}(E_{\gamma}) \simeq
\frac{1}{3\pi^2\hbar^2c^2}
\frac{\sigma_{\gamma n}(E_{\gamma})}
{E_{\gamma}} .
\end{equation}

This relation provides the experimental constraint on the E1 strength in the photon-energy region covered by the present $^{169}\mathrm{Tm}(\gamma,n)^{168}\mathrm{Tm}$ measurement.
The E1 strength function used in all TALYS-2.0 calculations of this work (the neutron-spectrum averaging of Eq.~(\ref{eq:eff_avg}), the $^{168}$Tm$(n,\gamma)$ cross section and rate, and the stellar $^{169}$Tm$(\gamma,n)$ rate) is obtained from Eq.~(\ref{eq:slo}) with the parameters of the joint fit listed in Table~\ref{tab:fitpar}, including its third component, converted with the relation above. The SLEGS-only parameters and those of Ref.~\cite{Gheorghe2026} are listed for comparison.

\begin{figure}[htbp]
\centering
\includegraphics[width=\linewidth]{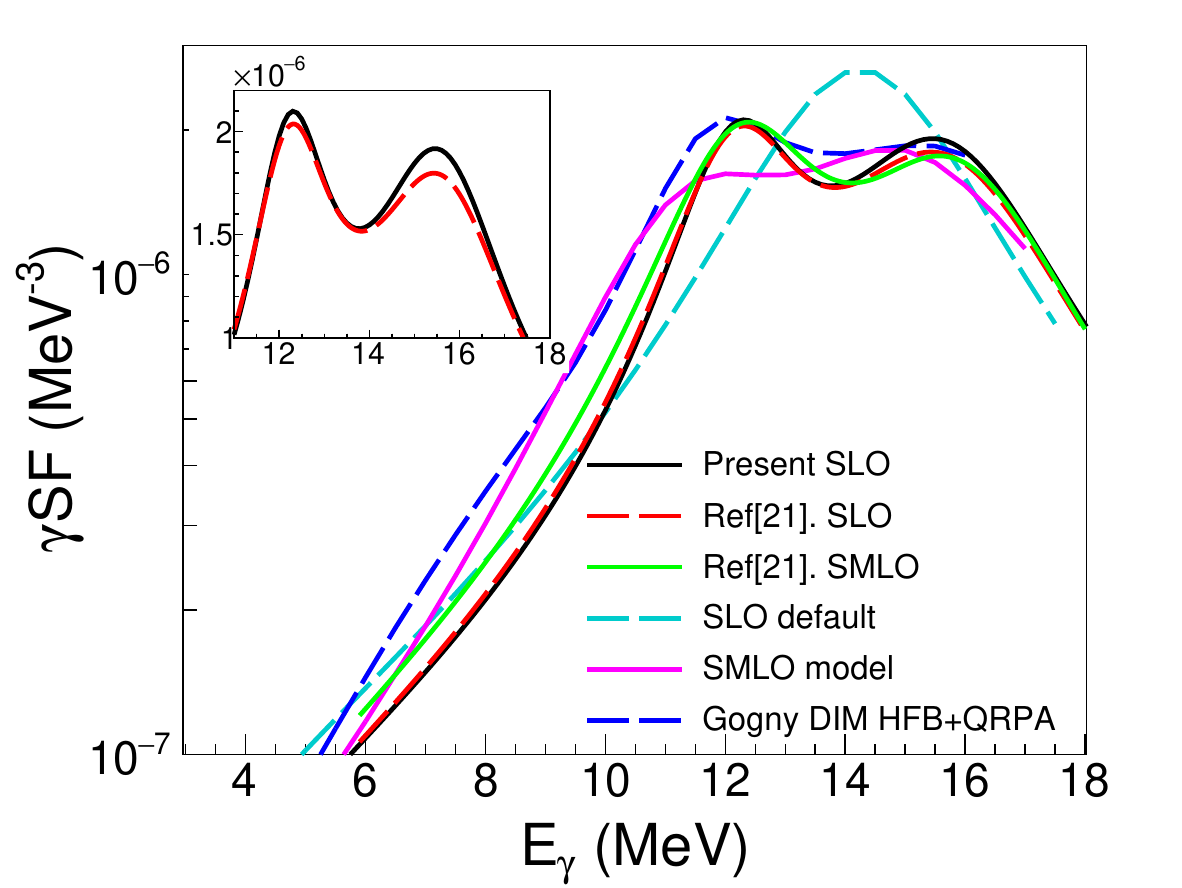}
\caption{Comparison of the experimental constraint $\gamma$SF of $^{169}\mathrm{Tm}$ with the theoretical E1 strength functions implemented in TALYS-2.0 (including SLO, SMLO, and Gogny D1M HFB+QRPA models), as well as the SLO and SMLO fitting results from Ref.~\cite{Gheorghe2026}. The inset compares the present SLO-constrained $\gamma$SF (solid) with the SLO fit of Ref.~\cite{Gheorghe2026} (dashed) on a linear scale.}
\label{fig:gsf_compare}
\end{figure}

Figure~\ref{fig:gsf_compare} compares the continuous $\gamma$SF extracted from SLO fits to the measured $^{169}\mathrm{Tm}(\gamma,n)^{168}\mathrm{Tm}$ cross section with several theoretical E1 strength prescriptions implemented in TALYS-2.0, including the standard Lorentzian (SLO), simplified modified Lorentzian (SMLO), and Gogny D1M HFB+QRPA models~\cite{GognyD1M,SMLO,Goriely2004}.
The figure also shows comparisons with the SLO and SMLO model fits reported in Ref.~\cite{Gheorghe2026}.
The Brink--Axel SLO prescription produces a smooth and broad strength distribution and does not reproduce the two-component GDR structure described by the present SLO fit to the measured excitation function.
The inset of Fig.~\ref{fig:gsf_compare} compares the present SLO-constrained strength with the SLO fit of Ref.~\cite{Gheorghe2026} on a linear scale: the two curves agree closely over the GDR region, differing by less than about 10\% at the peak maxima, which shows that the two SLO descriptions of the $^{169}$Tm photoneutron data yield essentially the same E1 strength function

In the lower-energy part of the measured region, around $E_\gamma \simeq 9$--11~MeV, the TALYS SMLO model follows the experimentally constrained strength relatively well, while the Gogny D1M HFB+QRPA calculation lies systematically above it by up to about a factor of two, although neither reproduces the shape of the split GDR over the full resonance region.
None of the prescriptions tested here reproduces both the low-energy slope and the split GDR; the E1 strength of the joint SLO fit is therefore used as the E1 input to the HF calculations, with its extrapolation below the measured energy range given by the fitted SLO shape.

\subsection{$^{168}$Tm$(n,\gamma)^{169}$Tm reaction cross-section}

The E1 $\gamma$SF extracted from the $^{169}$Tm$(\gamma,n)$ data also
constrains the inverse radiative capture reaction
$^{168}$Tm$(n,\gamma)^{169}$Tm.  Because $^{168}$Tm is unstable
($T_{1/2}\approx 93$~d), a direct measurement of neutron capture on it
is impractical, and evaluated cross sections rely almost entirely on
Hauser--Feshbach calculations with unconstrained $\gamma$SF prescriptions
\cite{TALYS,Hauser1952}.  By detailed balance,
the same E1 strength governs both directions, so the photoneutron
measurement provides an experimental anchor for the capture channel.

\begin{figure}[htbp]
\centering
\includegraphics[width=0.98\linewidth]{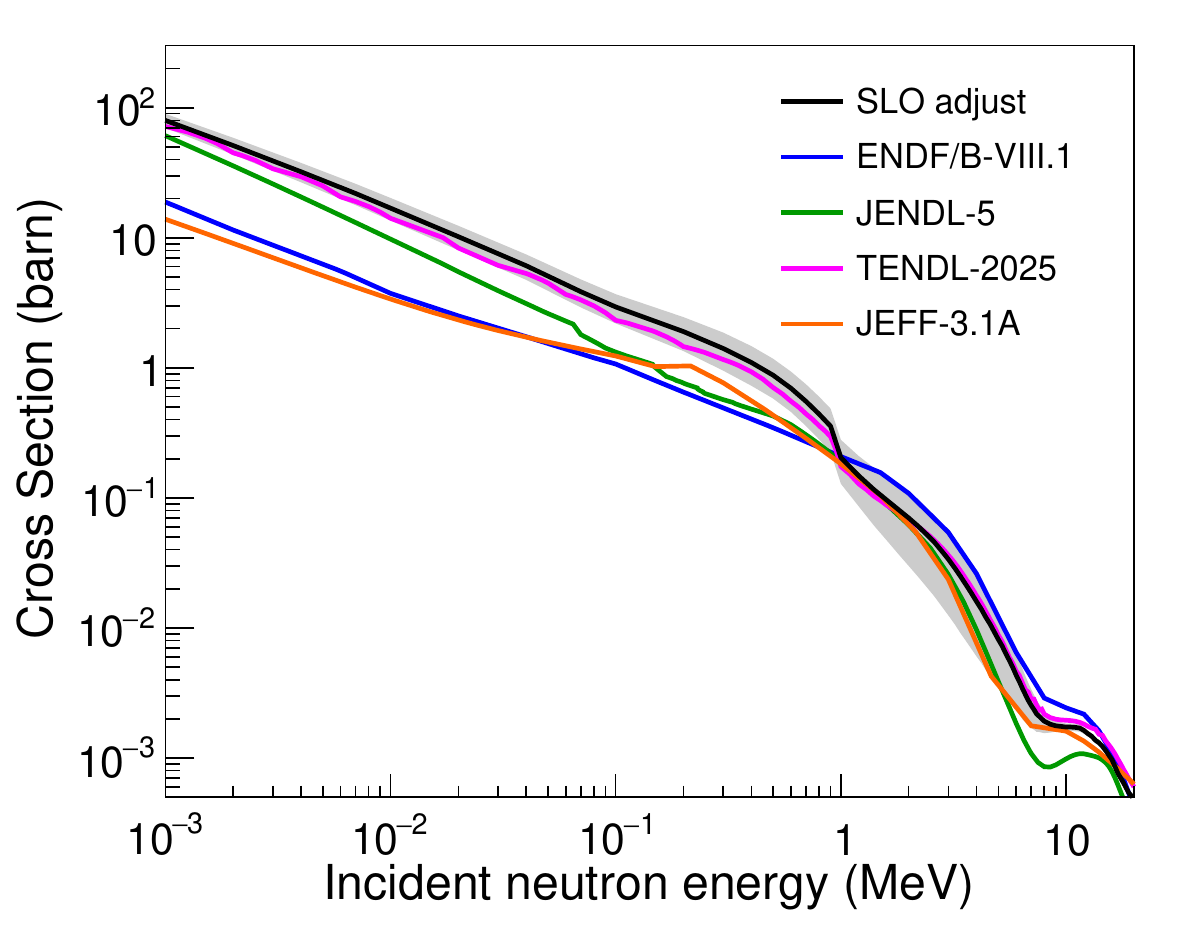}
\caption{Calculated $^{168}$Tm$(n,\gamma)^{169}$Tm cross sections using the experimentally constrained E1 $\gamma$SF (SLO adjust, black solid curve; the gray band shows the uncertainty from varying the nuclear level density model potential), compared with the ENDF/B-VIII.1, JENDL-5, TENDL-2025, and JEFF-3.1A evaluations.}
\label{fig:ngamma}
\end{figure}

In Fig.~\ref{fig:ngamma}, around $E_n \sim 1$~MeV all five calculations coincide within about a factor of two, but in the keV region relevant for the $s$ process the spread is considerably larger: the present SLO-constrained result lies above all four evaluations, exceeding ENDF/B-VIII.1 and JEFF-3.1A by a factor of a few at 1~keV, while JENDL-5 and TENDL-2025 lie closest to the constrained calculation.
This pattern points to the low-energy tail of the E1 $\gamma$SF as the main source of the discrepancy: in the evaluations this tail is purely a model extrapolation, whereas here it is anchored to the measured GDR strength through the Brink--Axel form.
However, the $\gamma$SF below $S_n$ remains an extrapolation of the fitted SLO shape and is not directly constrained by the present measurement, so the keV capture cross section still carries a model dependence, as partly reflected in the gray uncertainty band obtained by varying the nuclear level density model potential.
Above about 2~MeV the predictions diverge again, with ENDF/B-VIII.1 giving the largest cross section and JENDL-5 the smallest, so that the spread among the evaluations reaches up to about an order of magnitude.
Because the calculated keV capture cross section also depends on the chosen level density and on the neutron optical potential, neither of which is constrained by the present data, the differences from the evaluations at keV energies reflect this model dependence rather than a measurement-constrained result.

\section{Stellar reaction rate}
\label{sec:stellar_rate}

Laboratory photonuclear measurements directly constrain reactions on target nuclei in their ground states and therefore provide access to the ground-state photodisintegration rate.
Early work demonstrated that ground-state $(\gamma,n)$ rates can be determined experimentally by simulating the stellar photon bath with bremsstrahlung~\citep{Mohr2000} and through direct rate measurements~\citep{Vogt2001}. Subsequent comparisons of ground-state and stellar rates showed that the latter can far exceed the former when contributions from thermally populated target states are included~\citep{Vogt2001,Utsunomiya2006NPA}.
For the $^{169}\mathrm{Tm}(\gamma,n)^{168}\mathrm{Tm}$ reaction, the ground-state photodisintegration decay constant is obtained by folding the corresponding cross section with the thermal photon distribution,
\begin{equation}
\lambda_{\gamma n}^{\rm g.s.}(T)
=
\int_{S_n}^{\infty}
c\,n_{\gamma}(E_{\gamma},T)
\sigma_{\gamma n}^{\rm g.s.}(E_{\gamma})
\,dE_{\gamma},
\label{eq:lambda_gs}
\end{equation}
where $c$ is the speed of light and $n_{\gamma}(E_{\gamma},T)$ is the blackbody photon number density per unit energy,
\begin{equation}
n_{\gamma}(E_{\gamma},T)
=
\frac{8\pi}{h^{3}c^{3}}
\frac{E_{\gamma}^{2}}
{\exp(E_{\gamma}/k_{\rm B}T)-1}.
\label{eq:photon_density}
\end{equation}

For the ground-state reaction, the neutron separation energy, $S_n=8.03$~MeV, defines the reaction threshold.
When $E_{\gamma}$ is expressed in MeV and $\sigma_{\gamma n}$ in mb, the decay constant can be written as
\begin{equation}
\lambda_{\gamma n}^{\rm g.s.}(T_9)
=
3.953\times10^{14}
\int_{S_n}^{\infty}
\frac{E_{\gamma}^{2}
\sigma_{\gamma n}^{\rm g.s.}(E_{\gamma})}
{\exp(11.605E_{\gamma}/T_9)-1}
\,dE_{\gamma},
\label{eq:lambda_gs_numeric}
\end{equation}
where $\lambda_{\gamma n}^{\rm g.s.}$ is expressed in s$^{-1}$ and $T_9$ denotes the temperature in units of $10^{9}$~K.
Equations~(\ref{eq:lambda_gs})--(\ref{eq:lambda_gs_numeric}) define the ground-state rate for a given cross section.

Under stellar conditions, however, $^{169}\mathrm{Tm}$ nuclei are not restricted to their ground state.
At characteristic $p$-process temperatures, $T_9\simeq2$--$3.5$, corresponding to $k_{\rm B}T\simeq0.172$--$0.302$~MeV, low-lying excited states can be thermally populated and contribute to the photodisintegration rate~\citep{Mohr2006,Mohr2007}.
The stellar photodisintegration rate is obtained by thermally averaging over the ground and excited target states,
\begin{equation}
\lambda_{\gamma n}^{*}(T)
=
\frac{1}{G(T)}
\sum_i
\frac{2J_i+1}{2J_0+1}
\exp\left(-\frac{E_i}{k_{\rm B}T}\right)
\lambda_{\gamma n}^{(i)}(T),
\label{eq:stellar_rate}
\end{equation}
where $E_i$ and $J_i$ are the excitation energy and spin of the $i$th target state, respectively, $J_0$ is the ground-state spin, and $\lambda_{\gamma n}^{(i)}(T)$ is the photodisintegration rate from that state.
The temperature-dependent partition function $G(T)$, normalized to the ground-state statistical weight, is given by~\citep{Rauscher2000}
\begin{equation}
\begin{split}
(2J_0+1)G(T)
&= \sum_{i=0}^{i_m} (2J_i+1)e^{-E_i/k_{\rm B}T} \\
&\quad + \int_{E_{i_m}}^{E_{\max}}
\sum_{J,\pi} (2J+1)e^{-\epsilon/k_{\rm B}T}
\rho(\epsilon,J,\pi)\,d\epsilon ,
\end{split}
\label{eq:partition}
\end{equation}
where $\rho(\epsilon,J,\pi)$ is the spin- and parity-dependent nuclear level density.
In the present work, the stellar reaction rates $\lambda_{\gamma n}^{*}(T)$ were calculated with TALYS-2.0 using the $E1$ strength function obtained from the SLO fit to the experimental data as input (See Sec.~\ref{sec:gsf}).

\begin{figure}[htbp]
\centering
\includegraphics[width=\linewidth]{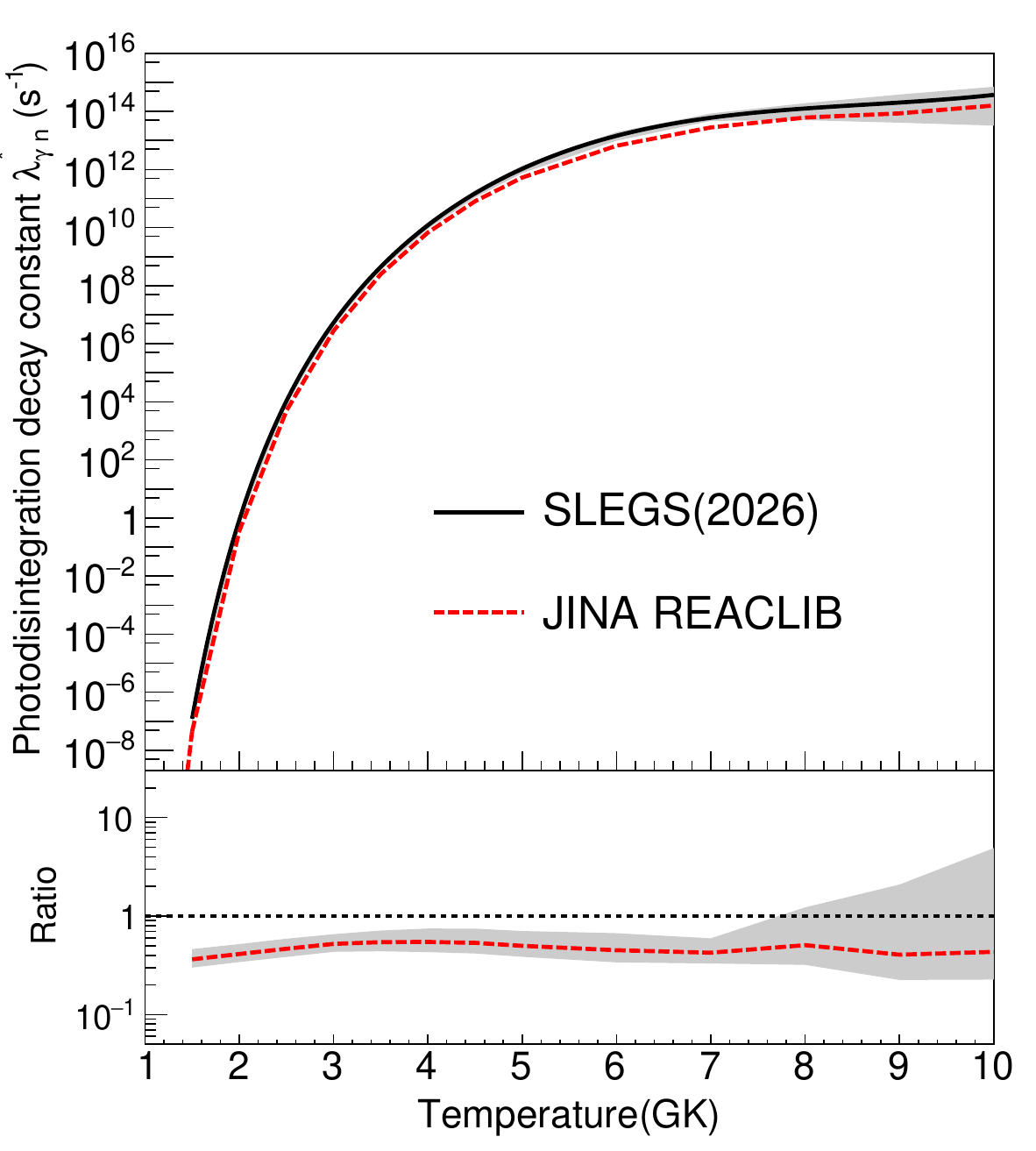}
\caption{Stellar $^{169}\mathrm{Tm}(\gamma,n)^{168}\mathrm{Tm}$ photodisintegration rate as a function of temperature $T_9$.
Upper panel: the black solid curve represents the TALYS-2.0 stellar rates calculated using the experimentally constrained $E1$ strength function over $T_9=1.5$--10.
The shaded band shows the uncertainty associated with the adopted NLD models, with bounds $\lambda^{*}\pm\Delta\lambda^{*}$ listed in Table~\ref{tab:rate}.
the red dashed curve is the JINA REACLIB rate; the REACLIB fit function was used with the required multiplication by $G_{^{168}\mathrm{Tm}}/G_{^{169}\mathrm{Tm}}$.
Lower panel: ratio of the JINA REACLIB rate to the present stellar rate.
The shaded band shows the uncertainty associated with the adopted NLD models.}
\label{fig:lambda}
\end{figure}

Figure~\ref{fig:lambda} compares the present stellar $^{169}\mathrm{Tm}(\gamma,n)^{168}\mathrm{Tm}$ rate, calculated using the experimentally constrained $\gamma$SF, with the JINA REACLIB database~\citep{Cyburt2010}.
Over the characteristic $p$-process temperature range, $T_9\simeq2$--3.5, the ratio of the JINA REACLIB rate to the present rate is approximately 0.44--0.55, corresponding to a present rate about 1.8--2.3 times higher.
The two rates exhibit broadly similar temperature dependences over $T_9=1.5$--10.
For $T_9\lesssim7$, the JINA REACLIB rate lies below the lower bound of the present NLD uncertainty band, whose relative half-width is approximately 20\%--33\%.
At $T_9=7$--10, this relative half-width increases to approximately 59\%--91\%, and the JINA REACLIB rate lies within the broadened band.
The present stellar rates are listed in Table~\ref{tab:rate}.

\begin{figure}[htbp]
\centering
\includegraphics[width=\columnwidth]{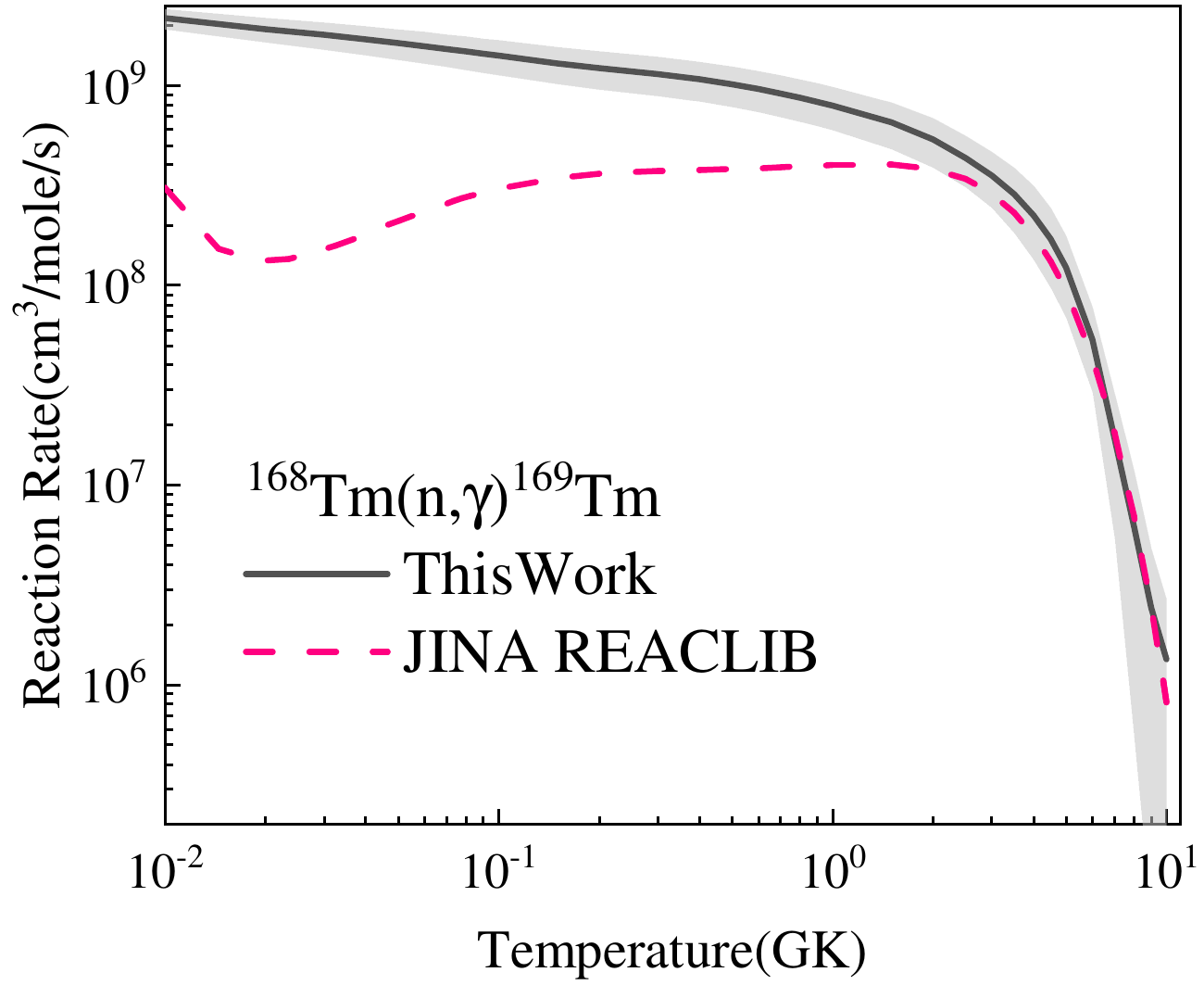}
\caption{Stellar $^{168}$Tm$(n,\gamma)^{169}$Tm reaction rate as a
function of temperature. Black solid curve (gray band): TALYS-2.0
calculation with the $\gamma$SF constrained in this work; the band shows the uncertainties from the nuclear level density models.
Magenta dashed curve: JINA REACLIB rate.}
\label{fig:ngamma_rate}
\end{figure}

The stellar reaction rate of $^{168}$Tm$(n,\gamma)^{169}$Tm was also predicted using TALYS-2.0, with the same model parameters as those employed in the $^{169}$Tm$(\gamma,n)^{168}$Tm rate calculations. Comparisons between the present rates and the JINA REACLIB rates are shown in Fig.~\ref{fig:ngamma_rate}. The JINA REACLIB rates significantly underestimate the present calculation by about a factor of  $4$ at $T_9\simeq0.1$.
The $^{168}$Tm$(n,\gamma)$ rate for $T_9=0.15$--$10$ is also listed in Table~\ref{tab:rate}.

\begin{table}[htbp]
\centering
\caption{The stellar reaction rates from the present TALYS-2.0 calculations:
the $^{169}\mathrm{Tm}(\gamma,n)^{168}\mathrm{Tm}$ photodisintegration rate
$\lambda_{\gamma n}^{*}$ (s$^{-1}$), the laboratory rate
$\lambda_{\gamma n}^{\mathrm{lab}}$ (s$^{-1}$) obtained by integrating the
three-component SLO fit of Table~\ref{tab:fitpar} using
Eq.~(\ref{eq:lambda_gs_numeric}), and the
$^{168}\mathrm{Tm}(n,\gamma)^{169}\mathrm{Tm}$ rate
$N_A\langle\sigma v\rangle^{*}$ (cm$^{3}$ mol$^{-1}$ s$^{-1}$) together with
its ground-state counterpart $N_A\langle\sigma v\rangle$
(cm$^{3}$ mol$^{-1}$ s$^{-1}$).
The stellar $(\gamma,n)$ rates at $T_9=0.15$--1.0 were obtained with a
modified version of TALYS-2.0 (see text), in which the rate integral is
evaluated in double precision with the $E_{\gamma}/k_{\rm B}T$ cutoff raised
from 80 to 700.
}

\label{tab:rate}
\begin{tabular}{lllll}
\hline\hline
$T_9$ & $\lambda_{\gamma n}^{*}$  & $\lambda_{\gamma n}^{\mathrm{lab}}$ & $N_A\langle\sigma v\rangle^{*}$  & $N_A\langle\sigma v\rangle$  \\
\midrule
0.15 & $6.39\times10^{-252}$ & $1.04\times10^{-254}$ & $1.26\times10^{9}$ & $1.31\times10^{9}$ \\
0.2 & $2.71\times10^{-184}$ & $3.95\times10^{-187}$ & $1.19\times10^{9}$ & $1.23\times10^{9}$ \\
0.5 & $2.98\times10^{-62}$ & $2.61\times10^{-65}$ & $9.80\times10^{8}$ & $9.91\times10^{8}$ \\
1.0 & $2.48\times10^{-21}$ & $1.60\times10^{-24}$ & $7.55\times10^{8}$ & $8.32\times10^{8}$ \\
1.5 & $1.21\times10^{-7}$ & $7.67\times10^{-11}$ & $6.20\times10^{8}$ & $7.47\times10^{8}$ \\
2.0 & $8.47\times10^{-1}$ & $5.90\times10^{-4}$ & $5.16\times10^{8}$ & $6.85\times10^{8}$ \\
2.5 & $1.06\times10^{4}$ & $8.54\times10^{0}$ & $4.24\times10^{8}$ & $6.35\times10^{8}$ \\
3.0 & $5.52\times10^{6}$ & $5.32\times10^{3}$ & $3.42\times10^{8}$ & $5.92\times10^{8}$ \\
3.5 & $4.62\times10^{8}$ & $5.48\times10^{5}$ & $2.69\times10^{8}$ & $5.53\times10^{8}$ \\
4.0 & $1.22\times10^{10}$ & $1.83\times10^{7}$ & $2.07\times10^{8}$ & $5.19\times10^{8}$ \\
5.0 & $9.85\times10^{11}$ & $2.67\times10^{9}$ & $1.10\times10^{8}$ & $4.61\times10^{8}$ \\
6.0 & $1.38\times10^{13}$ & $8.01\times10^{10}$ & $4.88\times10^{7}$ & $4.13\times10^{8}$ \\
8.0 & $1.61\times10^{14}$ & $6.74\times10^{12}$ & $7.19\times10^{6}$ & $3.40\times10^{8}$ \\
10.0 & $4.27\times10^{14}$ & $1.14\times10^{14}$ & $1.27\times10^{6}$ & $2.87\times10^{8}$ \\
\hline\hline
\end{tabular}
\end{table}

\section{Implications for the $\gamma$-Process calculations}
\label{sec:astro}

We use MESA (v12778;\cite{MESA2011,MESA2015,MESA2018,MESA2023}) to evolve a  $20\,M_\odot$, Z = 0.02 star from the ZAMS until the infall velocity reaches $1000\,\mathrm{km\,s^{-1}}$ in the iron core ~\cite{Xin2025,Imasheva2023}. The CCSN shock propagation and explosive nucleosynthesis are modeled with the thermal-bomb prescription of Ref.~\cite{Imasheva2023}.  The mass cut is placed at $S=4\,k_{\rm B}\,\mathrm{baryon}^{-1}$ \cite{Heger2010}, with the interior material removed to form the compact remnant. The explosion is initiated by depositing energy into a 0.01 $M_\odot$ shell to yield a total energy of $1.0\times10^{51}$~erg. Using the resulting $20\,M_\odot$ CCSN model, we perform post-processing p-process calculations along selected mass-zone trajectories, following Ref.~\cite{Rapp2006}. The network (7852 nuclides, 48,965 reactions, H to Cn) is solved with NucNet Tools\cite{Meyer2012}. Initial abundances and thermodynamic trajectories are taken from the MESA and thermal-bomb calculations above. Reaction rates are adopted from the JINA REACLIB database\cite{Cyburt2010}.

As the supernova shock propagates through the O/Ne shell, it undergoes a rapid increase in temperature and density, followed by an approximately exponential decline after shock passage. During the brief high-temperature phase, the intense photon field activates photodisintegration reactions, predominantly $(\gamma,n)$ together with $(\gamma,p)$ and $(\gamma,\alpha)$ channels, and drives the initial seed nuclei toward the proton-rich side of the valley of stability. After shock passage, the temperature and density decrease rapidly, photodisintegration reactions freeze out, and the unstable proton-rich nuclei subsequently decay toward stability, establishing the post $p$-process abundance pattern. We chose the zone at $M=3.25\,M_\odot$, which reaches a moderate $T_{9,\rm max}\approx2.5$, so that the sensitivity of the local reaction flow to the $^{169}$Tm$(\gamma,n)$ rate is expected to be most pronounced. After the shock, the network calculation is continued for an additional 10~Myr at low temperature and density to allow the unstable nuclei to decay toward stability and form the final abundances.

\begin{figure}[htbp]
\centering
\includegraphics[width=\linewidth]{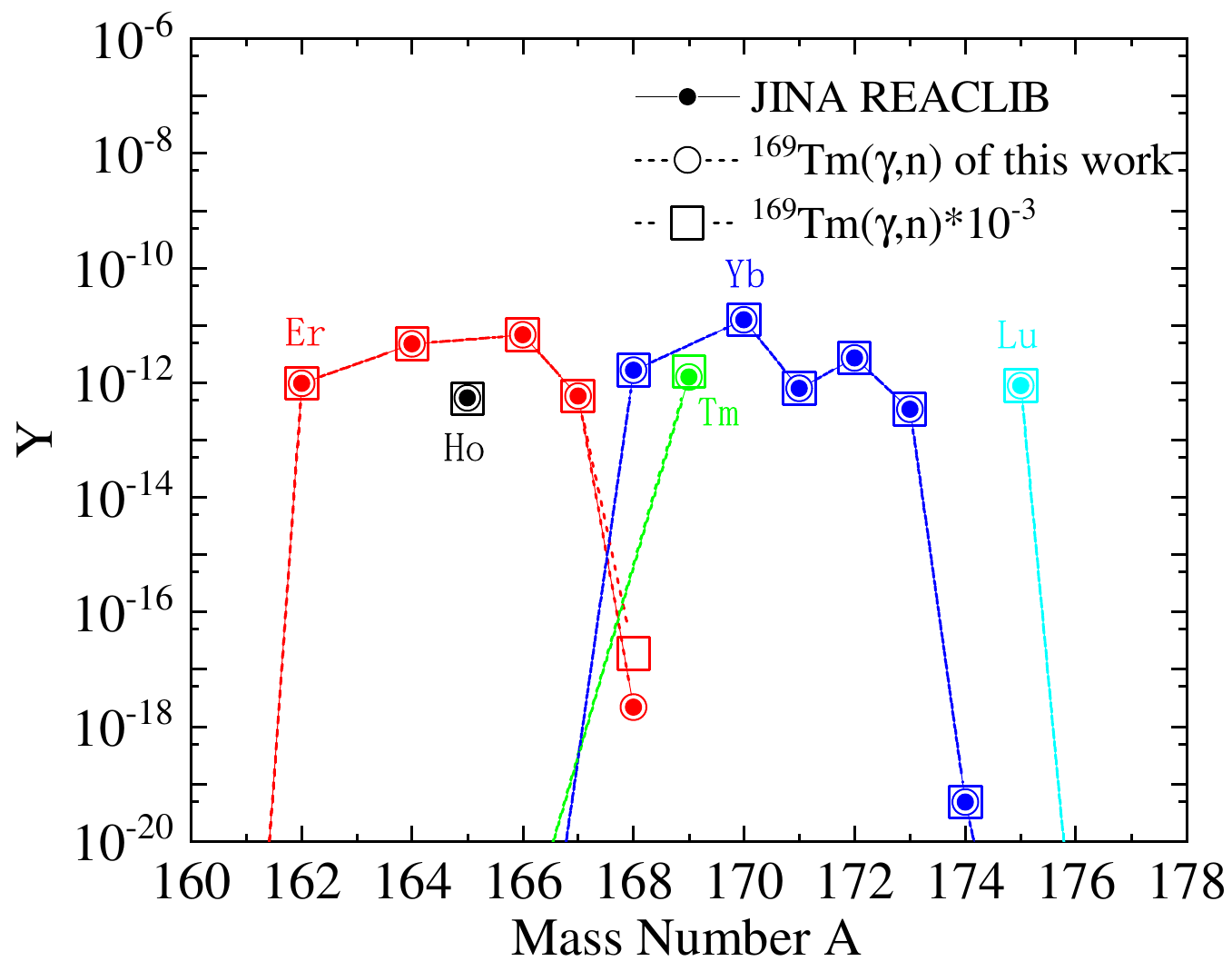}
\caption{Final abundances in the rare-earth region around $A\simeq169$ after the decay phase for the $M=3.25\,M_\odot$ trajectory. Solid curves with filled circles: default JINA REACLIB $^{169}$Tm$(\gamma,n)$ rate \cite{Cyburt2010}; dotted curves with open circles represent the rate of this work; dashed curves with open squares: REACLIB rate reduced by a factor of $10^{3}$, an extreme variation included only to illustrate the sensitivity.}
\label{fig:abundance}
\end{figure}

Figure~\ref{fig:abundance} shows the final abundances $Y$ in the rare-earth mass region after the full 10~Myr decay phase.
The solid curves with filled circles correspond to the default JINA REACLIB $^{169}$Tm$(\gamma,n)$ rate, while the dotted curves with open circles show the results obtained with the present $^{169}$Tm$(\gamma,n)$ rate. The dashed curves with open squares show the result obtained when the rate is reduced by a factor of 10$^{3}$, which is included to make the sensitivity of the nucleosynthesis to this reaction rate more apparent.
As shown in Fig.~\ref{fig:lambda}, within the characteristic $p$-process temperature range the present rate is about a factor of two larger than the JINA REACLIB rate. Figure~\ref{fig:abundance} shows that a rate difference of this size has a negligible effect on the final abundances Y. Even in the more extreme case where the rate is reduced by a factor of 10³ relative to the JINA REACLIB rate, only the $^{168}$Er abundance increases by about one order of magnitude. The reason is that suppressing this channel leaves more $^{169}$Tm, and accordingly more $^{168}$Tm is retained relative to the default case; the decay product of $^{168}$Tm is $^{168}$Er. Despite the enhancement of $^{168}$Er, the overall Er abundance pattern is only weakly affected. In summary, while the effect of the $^{169}$Tm$(\gamma,n)$ reaction rate on the final $p$-process abundances is small, it does have some influence on the distribution of the local reaction flow among the competing channels in the Tm–Er region.

\section{Conclusion}
\label{sec:conclusion}

The $^{169}\mathrm{Tm}(\gamma,n)^{168}\mathrm{Tm}$ cross section was measured at SLEGS with quasi-monoenergetic LCS $\gamma$-ray beams from 8.27~MeV to 14.61~MeV. Monoenergetic cross sections at 18 energies were obtained by iterative unfolding, with total uncertainties of about 2\%--7\%. 
Below 12~MeV the present cross sections agree well with the analyzed NewSUBARU data and with the IAEA-2019. In the region of the first GDR component at 12--14~MeV, the present cross sections are larger than the NewSUBARU data by up to about 10\%, which lies within the combined uncertainties. A joint fit to the present and NewSUBARU data constrains the GDR, with the first component at $E_{m1}\simeq12.3$~MeV and $\Gamma_1\simeq2.2$~MeV. The E1 strength function derived from it agrees with that obtained from the NewSUBARU fit to within about 10\% at the peak maxima.
With the experimentally constrained E1 strength, the stellar $^{169}\mathrm{Tm}(\gamma,n)$ rate and the $^{168}\mathrm{Tm}(n,\gamma)^{169}\mathrm{Tm}$ cross section and stellar rate were calculated in the Hauser--Feshbach framework. At $T_9=2$--3.5 the photodisintegration rate is larger than the JINA REACLIB rate by a factor of about 2, with a remaining uncertainty from the nuclear level density. Network calculations for the O/Ne shell of a $20\,M_\odot$ core-collapse supernova model show that the revised rate has a negligible effect on the final $p$-nucleus abundances; even when the rate is reduced by a factor of $10^{3}$, only the $^{168}$Er abundance changes appreciably.

\begin{acknowledgments}
This work was supported by the National Natural Science Foundation of China (Grants No.~12465024 and No.~12365018), the Natural Science Foundation of Inner Mongolia (Grants No.~2024ZD23), the Central Government Guidance for Local Science and Technology Development Funds Project (Grant No.~2025ZY0067), and Hungarian National Research, Development and Innovation Fund NKFIH (K147010).
The authors gratefully acknowledge the accelerator operators and beamline engineers at the SLEGS facility~\cite{SLGES} of the SSRF~\cite{SSRF} for delivering a stable $\gamma$-ray beam throughout the experiment and for their invaluable technical assistance. We thank I.~Gheorghe for providing the analyzed NewSUBARU $^{169}$Tm photoneutron data (private communication).

\end{acknowledgments}

\section*{Data availability}
The cross-section data of Table~\ref{tab:cross_section} and the stellar rates of Table~\ref{tab:rate} are available from the corresponding authors upon reasonable request. The NewSUBARU $^{169}$Tm photoneutron data used in this work were provided by I.~Gheorghe (private communication); the analysis procedure is described in Ref.~\cite{Gheorghe2026}.

\bibliographystyle{apsrev4-2}
\bibliography{Tm169_PRC}

\end{document}